\documentclass[twocolumn,showpacs,amssymb,prl]{revtex4}

\usepackage{graphicx}
\usepackage{dcolumn}
\usepackage{bm}

\begin{document}

\preprint{APS/123-QED}

\title{Recovering Metallicity in A$_4$C$_{60}$: The Case of Monomeric  Li$_4$C$_{60}$}

\author{M.~Ricc\`o}%
\email{Mauro.Ricco@fis.unipr.it}
\author{M.~Belli}%
\author{D.~Pontiroli}%
\author{M.~Mazzani}%
\author{T.~Shiroka}%
\affiliation{Dipartimento di Fisica and 
Consorzio Nazionale Interuniversitario per le Scienze Fisiche della Materia,
Universit\`a di Parma, Via G.~Usberti 7/a, 43100 Parma, Italy}
\author{D.~Ar\v{c}on}%
\author{A. Zorko}
\affiliation{Solid State Physics Department, Jozef Stefan Institute, Jamova 39, SI-1000, 
Ljubljana, Slovenia}
\author{S.~Margadonna}%
\affiliation{School of Chemistry, University of Edinburgh, West Mains Road, 
EH9 3JJ Edinburgh, UK}
\author{G.~Ruani}%
\affiliation{ISMN-CNR, Via P.\ Gobetti 101, 40129 Bologna, Italy}
\date{\today}%

\begin{abstract}
The restoration of metallicity in the high-temperature, cubic phase of
Li$_4$C$_{60}$ represents a remarkable feature for a member of the 
A$_4$C$_{60}$ family (A = alkali metal), invariably found to be insulators.
Structural and resonance technique investigations on Li$_4$C$_{60}$ 
at $T$ > 600 K, show that its fcc structure is associated with a complete 
(4$e^-$) charge transfer to C$_{60}$ and a sparsely populated Fermi level.
These findings not only emphasize the crucial role played by lattice 
symmetry in fulleride transport properties, but also re-dimension the 
role of Jahn-Teller effects in band structure determination. 
\end{abstract}


\maketitle
In recent years the competition between Coulomb repulsion, kinetic energy, Jahn-Teller 
effect, and Hund's coupling rules has been subject of intensive research in the
domain of alkali-doped fullerenes.
Apparently, most of the phenomena, including high-$T_c$ superconductivity in 
A$_3$C$_{60}$, seem to be relatively well understood.
Nevertheless, the reason why A$_{4}$C$_{60}$ are in general nonmagnetic insulators 
\cite{Murphy92,Lukyanchuk95} is still puzzling and remains subject of strong controversy. 
The situation is made even more complex as, differently from fcc A$_3$C$_{60}$, the 
A$_{4}$C$_{60}$ arising from large-size alkali metals (A = K, Rb, Cs) adopt a 
bct structure \cite{Fleming91}.
Hence, the natural suggestion that the difference in lattice structure
favors the metallicity in the former and an insulating behavior in the latter, 
with the coupling to $H_{\mathrm{g}}$ Jahn-Teller phonons accounting for the lack 
of magnetic ordering in A$_{4}$C$_{60}$\cite{Han00}.  

Doping the C$_{60}$ with the significantly smaller Li$^+$ and Na$^+$ alkali 
ions leads to interesting deviations from the standard behavior. 
Indeed, Li$_x$C$_{60}$ can easily form compounds with $x>6$ \cite{Tomaselli01}, 
whereas in Na$_{4}$C$_{60}$ the fullerene molecules form a 2D planar polymer, 
with the molecular units linked by four single C--C bonds \cite{Oszlanyi97}. 
As commonly observed in singly bonded C$_{60}$ polymers \cite{Bendele98},
the latter system too was reported to be metallic \cite{Oszlanyi97,Rezzouk02}.
Li$_4$C$_{60}$, though, represents a different situation. At normal conditions, it 
is a 2D polymer displaying a novel architecture based on the coexistence of single 
and double bonds oriented along two perpendicular directions within the polymer 
plane \cite{Margadonna04}. 
Although the presence of single bonds could suggest a metallic behavior, 
NMR, Raman and ESR spectroscopy show beyond doubt the diamagnetic and 
insulating character of this polymer \cite{Ricco05}. Important questions 
hence arise both regarding the role of structure, as well as of 
polymerization in depressing the electron delocalization 
and destroying the expected metallicity.

In this article we try to an\-swer these ques\-tions by a thorough 
investigation of the high temperature monomeric Li$_4$C$_{60}$ phase, 
obtained by thermally induced depolymerization which takes place above 590K.
First we show that, in spite of its stoichiometry, the Li$_4$C$_{60}$ monomeric 
phase retains a closed packed (fcc) cubic structure, and therefore can be directly 
compared to other, well known A$_3$C$_{60}$ fcc systems. Subsequently, as revealed 
from the ESR lineshape, $^{7}$Li and $^{13}$C NMR Knight shift and Raman bandwidth, 
we demonstrate the metallicity of the monomeric Li$_4$C$_{60}$. Finally, a full 
charge transfer of four electrons to C$_{60}$ is clearly evidenced by the Raman 
A$_g$(2) mode shift. These results not only challenge established theories on alkali 
intercalated fullerides, but also stimulate a re-consideration of the relative 
importance of the many competing mechanisms present in these systems.

\begin{figure}
\includegraphics[width=0.45\textwidth]{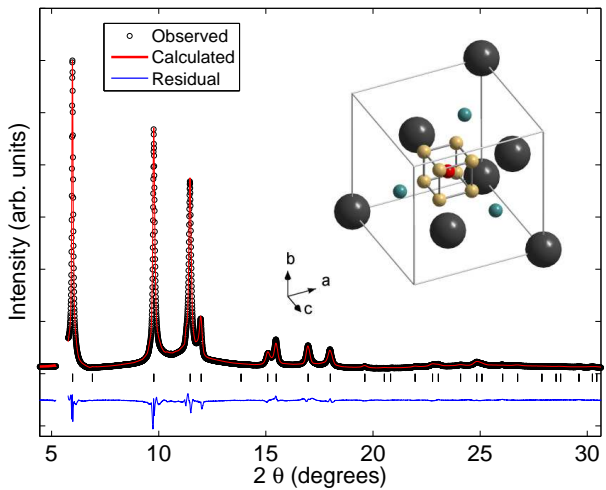}
\caption{\label{fig:Li4C60_XRD}(color online). 
Calculated and observed synchrotron powder diffraction profile 
of Li$_{4}$C$_{60}$ at 773 K, taken with $\lambda = 0.85055$ \AA\ 
($R_{\mathrm{wp}} = 4.04$\%, $R_{\mathrm{exp}} = 2.96$\%).
Inset: crystal structure of Li$_{4}$C$_{60}$ monomer. 
Lithium atoms (small spheres) occupy three different non-equivalent 
positions in the fcc lattice ($a = 14.122$ \AA)  
formed by fullerene units (large spheres).}	
\end{figure}

The Li$_{4}$C$_{60}$ samples were prepared following intercalation procedures 
described in detail elsewhere \cite{Ricco05}.
High-resolution synchrotron X-ray powder diffraction data on Li$_{4}$C$_{60}$ 
were collected on the ID31 beamline at the ESRF facility (Grenoble) at 773 K 
($\lambda = 0.85055$~\AA), well above the depolymerization temperature 
($T_{\mathrm{dep}} \sim 590$ K) (Fig.~\ref{fig:Li4C60_XRD}).
The rebinned data 
were then analyzed using the Fullprof suite of Rietveld analysis programs.

All the peaks of the 
powder diffraction profile were indexed with a 
face centered cubic cell, except for an evident shoulder at $2 \theta \sim 6^{\circ}$ 
on the leftmost peak, present also in pristine cubic C$_{60}$ and generally associated 
to stacking faults among the ($111$) planes \cite{Moret92}.

Starting with a face centered cubic structure (space group $F m \bar{3}m$), data analysis 
using the Le Bail pattern decomposition technique proceeded smoothly, resulting in 
a lattice parameter $a = 14.122$ \AA\ (agreement factors: 
$R_{\mathrm{wp}} = 3.33$\%, $R_{\mathrm{exp}} = 2.44$\%). 
It is worth noting that the $\sim 9.97$ \AA\ inter-ball 
distance between two nearest C$_{60}$ units, although slightly smaller than in
pristine C$_{60}$ (due to positive electrostatic pressure), is fully compatible with 
that in monomeric fullerenes.

To account for the orientational disorder in the high temperature phase, during 
Rietveld refinement the fullerene units were modeled with quasi-spherical shells 
centered at $4a$ ($0, 0, 0$) sites. Their scattering density was described 
in terms of symmetry-adapted spherical harmonics (SASH) which, 
compatibly with the symmetry of the system, comprise only $K_{l,j} (\theta, \phi)$ 
terms with $l = 0$, 6, 10. The respective coefficients $c_{l,j}$, treated 
as free refinement parameters, have fitted values $c_{6,1} = 0.035(3)$ and 
$c_{10,1} = -0.156(13)$. The latter is indicative of an excess electronic density 
along the $[1 1 1]$ directions, where tetrahedral lithium ions are expected to 
reside. Nevertheless, the relatively small value of the $c_{10,1}$ coefficient 
(to be compared with c$_{10,1} = -0.26(3)$ in Li$_{2}$CsC$_{60}$ \cite{Margadonna99a}) 
suggests a considerably weakened Li$^+$--C interaction.

While three of the intercalated lithium ions were easily localized in the 
octahedral (4$b$) and the two tetrahedral (8$c$) sites, the location of the 
fourth excess Li$^+$ proved a more difficult task. 
A difference Fourier analysis revealed the existence of scattered
intensity in the vicinity of the 32$f$ sites ($x,x,x$, with $x \sim 0.375$), 
which define a cube centered at the ($\frac{1}{2}, \frac{1}{2}, \frac{1}{2}$) 
cell position. 
Subsequent Rietveld refinements, repeated after introducing the missing 
Li$^+$ ion in these sites and allowing for its occupation number to 
vary, indicated that the excess alkali ion is randomly distributed over 
the corners of the cube with $\frac{1}{8}$ occupancy. 


The proposed structure (see inset in Fig.~\ref{fig:Li4C60_XRD}) 
shows deep analogies with another cubic fulleride, Li$_{3}$CsC$_{60}$ 
\cite{Margadonna99b}, which presents similar stoichiometry, 
but only a partial charge transfer of 3 electrons \cite{Kosaka99}.
The accurate evaluation of Li$^+$--C contact distances ($\sim 2.4$ \AA) 
confirmed once more their weak mutual interactions, strongly suggesting 
an almost \textit{complete} charge transfer from the intercalated metal 
to the C$_{60}$ units.
Hence, the structural analysis not only shows that the high-temperature 
treatment of the 
2D polymeric Li$_4$C$_{60}$ leads to a fcc monomeric phase, 
analogous to those observed in A$_3$C$_{60}$, but it gives also precise, 
though indirect, hints about the electronic properties, whose direct 
investigation is described in detail below.

\begin{figure}
\includegraphics[width=0.45\textwidth]{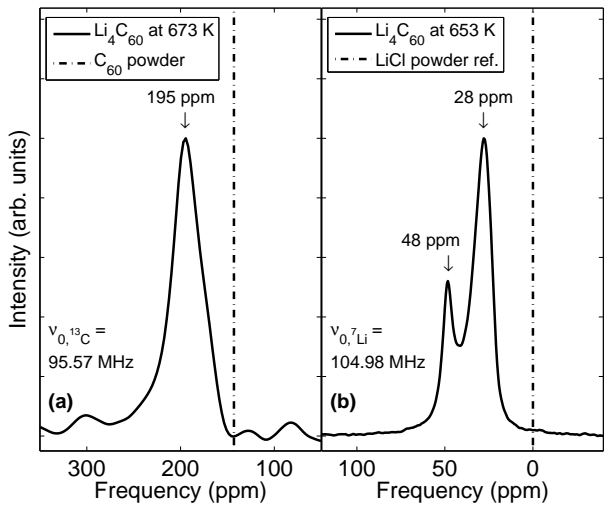}
\caption{\label{fig:Li4C60_NMR}NMR spectra of Li$_{4}$C$_{60}$ in 
the monomer phase: (a) $^{13}$C NMR spectrum at 
673 K referenced to tetramethyl silane. The pristine C$_{60}$ line position at +143 
ppm is also shown by dotted line; (b) $^{7}$Li spectrum at 653 K referenced to the narrow line of 
anhydrous LiCl powder measured at the same temperature.
}
\end{figure}

A straightforward evidence about the metallic character of the Li$_4$C$_{60}$ 
monomer is given by the presence of a Knight shift in Nuclear Magnetic Resonance 
(NMR) spectra of both $^{13}$C and $^{7}$Li nuclei.
The room temperature $^{13}$C NMR spectrum of the as-prepared Li$_4$C$_{60}$ 
polymer confirms the insulating nature of the compound \cite{Ricco05},
whereas heating the sample to high temperatures introduces significant spectral 
changes. The nearly structureless spectrum measured at $T = 673$ K 
(Fig.~\ref{fig:Li4C60_NMR}), well above the depolymerization transition, 
is fully compatible with freely rotating C$_{60}$ molecules, as expected in a
monomeric phase. 
What we find really surprising is the considerable positive shift of the 
$^{13}$C NMR line, $\delta\nu = 195$ ppm, much larger than typical values 
reported for metallic A$_3$C$_{60}$ fullerides \cite{Maniwa96}.
This fact clearly suggests an increased value of the isotropic Knight shift and 
could be rationalized in the same framework of the ESR results (\textit{vide infra}) 
by a substantial increase of the isotropic hyperfine coupling for the C$_{60}^{4-}$
state, as compared to the one measured for C$_{60}^{3-}$.

The $^{7}$Li NMR spectrum of the monomer phase of Li$_{4}$C$_{60}$, 
reported in Fig.~\ref{fig:Li4C60_NMR}(b), 
shows two partially merged peaks centered at +28 and +48 ppm relative to LiCl. 
They 
comprise all of the transitions of the $^7$Li nucleus
($I= \frac{3}{2}$), as confirmed by nutation angle analysis. 
The upshift can easily be attributed to a metallic behavior, since typical shifts 
in other, insulating lithium fullerides would fall in the few ppm 
range \cite{Tomaselli01,Ricco05,Cristofolini99}. 
It seems to follow quite well the common trend of metallic fullerides, 
where smaller alkali metals give rise to lower shift values \cite{Pennington96}.

Even though the non-negligible temperature dependence of the relative
intensities prevented us from a detailed analysis, we may still attribute
the more shifted peak to the disordered lithium ions in the $32f$ sites,
as a higher electronic density is expected at these asymmetric positions, 
due to a theoretically predicted enhancement of the fullerene--alkali metal 
electronic hopping \cite{Gunnarsson98}.

\begin{figure}
\includegraphics[width=0.45\textwidth]{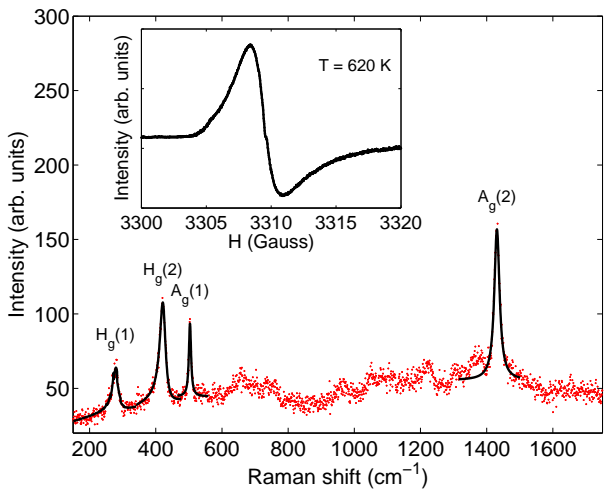}
\caption{\label{fig:Li4C60_Raman}(color online). 
Raman spectrum of monomeric Li$_{4}$C$_{60}$ (dots) taken at 648 K 
with a 1.96 eV excitation. Only four Raman modes, associated to A$_g$ 
and H$_g$ cage vibrations, are clearly detected and respectively fitted
to Lorentzian and BWF curves (solid lines). The H$_g$ resonances appear 
broadened and asymmetric, in agreement with the metallic nature of the compound.
Inset: The anisotropic Dysonian lineshape of the ESR signal at 620 K also 
supports the sample's metallicity.}
\end{figure}

Further evidence on the metallic ground state of the high-temperature phase comes from 
ESR experiments. At room temperature, the as-prepared polymeric Li$_4$C$_{60}$ displays 
an extremely narrow X-band ESR signal ($\Delta H_{pp}=0.27(3)$ G), described quite
well by a simple Lorentzian lineshape. 
The calibrated ESR signal intensity corresponds to a small spin susceptibility, 
$0.24\times 10^{-4}$ emu/mol, and to a $g$-factor insignificantly larger than 2. 
As the sample is heated, the ESR signal undergoes dramatic changes. Indeed, 
already at 370 K, the lineshape becomes slightly anisotropic, whereas for 
temperatures above 520 K the anisotropy becomes extremely pronounced 
(see inset in Fig.~\ref{fig:Li4C60_Raman}). The lineshape closely 
resembles a Dysonian function \cite{Dyson55}, providing a clearcut 
indication of the metallic character of the monomer.

In parallel with lineshape evolution, signal intensity too shows a sudden 
increase by nearly an order of magnitude and, at $T= 620$ K, the 
ESR spin susceptibility reaches $\chi_{S}=1.3(2) \cdot 10^{-4}$ emu/mol, a 
value comparable with those measured in monomeric metallic 
A$_x$C$_{60}$ \cite{Tanigaki95}. 
We would like to stress that, despite the increase, this value remains 
nevertheless smaller than any other known spin susceptibility in metallic 
C$_{60}$ phases. In fact, the measured Pauli susceptibility $\chi_S$ implies 
a density of states at the Fermi level $N(E_F)\sim 5\; {\rm states/eV}$, 
to be compared with  $N(E_F)\sim 14\; {\rm states/eV}$ obtained in a typical
A$_3$C$_{60}$ compound with similar lattice parameter \cite{Ramirez92}.

This reduction in the density of states could arise either from an increased bandwidth $W$, or 
from strong spin correlation effects.
An increase in bandwidth up to 40\% \cite{Gunnarsson98} is already 
expected in bct A$_{4}$C$_{60}$ systems, due to the presence of alkali
ions in non-symmetric positions. In our case, the same effect could 
arise from the Li ions located in the $32f$ sites.
If this were the case, not only the Pauli susceptibility would be reduced, 
but also the effective ESR $g$-factor would be affected.
As a comparison, typical $g$-factor values in Rb$_3$C$_{60}$ and 
K$_3$C$_{60}$ are 1.9945 and 2.0003 respectively \cite{Tanigaki95}, whereas
in the high-temperature fcc phase of Li$_4$C$_{60}$ we find $g=2.0036(5)$, 
a value strikingly different from the others. Such an anomalously high 
$g$-factor could reflect either an enhanced C$_{60}$--Li electron hopping 
mechanism or continuous changes in the spin-orbit scattering due to 
lithium disorder in Li$_4$C$_{60}$.

Raman spectra were recorded in a back-scattering geometry from room 
temperature up to 648 K. A relatively low laser intensity (200 W/cm$^2$) 
helped us to avoid the photo-damage of the investigated samples.
The room temperature Raman scattering is characterized by a large 
number of narrow lines, typical of an insulating, low-symmetry 
polymer \cite{Ricco05}. 
On the other hand, at high temperatures most of the peaks disappear and only four of them  
(Fig.~\ref{fig:Li4C60_Raman}), attributed to the H$_g$(1), H$_g$(2), A$_g$(1) 
and A$_g$(2) modes, could be observed.
While the A$_g$ modes are symmetric and could be fitted to Lorentzian peaks, 
the H$_g$ modes display an asymmetric shape and were thus analyzed in terms of 
Breit-Wigner-Fano (BWF) resonances \cite{Zhou93}; the best fit values are 
reported in Tab.~\ref{tab:Raman_data}.
The spectrum is very similar to the one observed in A$_{3}$C$_{60}$ (A = K, Rb) 
(\cite{Dresselhaus96} and refs.\ therein), once more confirming the metallic 
nature of the compound.

\begin{table}
\caption{\label{tab:Raman_data}Fitted parameters for Li$_4$C$_{60}$ 
and pristine C$_{60}$ Raman spectra taken at 648 K. 
The line shifts $\omega_0$ and widths $\Gamma$ are expressed 
in cm$^{-1}$, the Fano parameter $q$ is adimensional.}
\begin{ruledtabular}
\begin{tabular}{rcrrrr}
                &            & H$_g$(1)~~ & H$_g$(2)~~ & A$_g$(1)~~ & A$_g$(2)~ \\ 
\hline
Li$_4$C$_{60}$: & $\omega_0$ & 279.5(7) & 422.5(4) & 503.3(2)  & 1431.1(2)  \\
                & $\Gamma$   & 21(1)    & 21.0(6)  & 8.0(5)    & 19.2(5)    \\
                & $q$        & $-8(2)$  & $-7(1)$  & ---       & ---        \\
     C$_{60}$:  & $\omega_0$ & 275.9(1) & ---\footnotemark[1]  & 497.6(1)   & 1458.8(2) \\
                & $\Gamma$   & 5.8(5)   & ---\footnotemark[1]  & 3.5(3)     & 5.6(7)    \\
\end{tabular}
\end{ruledtabular}
\footnotetext[1]{H$_{\mathrm{g}}$(2) mode in pristine C$_{60}$ not detectable at this temperature.}
\end{table}

To determine the charge transfer towards C$_{60}$ we use the well known 
fact that A$_g$(2) mode energy shifts are proportional to the number of 
transferred electrons 
(6--7 cm$^{-1}$/electron \cite{Wagberg02}).
However, to discriminate pure\-ly thermal effects from alkali doping, we perform 
(at the same temperature) a comparative Raman scattering in undoped C$_{60}$,
whose modes could be fitted with Lorentzian curves (see 
Tab.~\ref{tab:Raman_data}).

The doping-related energy shift of the A$_g$(2) mode, 27.8 cm$^{-1}$,
is consistent with a charge transfer of four electrons per C$_{60}$ 
(6.9 cm$^{-1}$/electron).
Finally, the presence of BWF line shapes indicates an interaction 
of the H$_g$ modes with free electrons in the $t_{1u}$-derived state 
of C$_{60}$ molecule. From the phonon frequency $\omega_i$ and 
the full width at half maximum $\Gamma_i$ of the $i^{\mathrm{th}}$ 
mode, one can estimate the electron-phonon coupling constants $\lambda_i$ 
(\cite{Winter96} and refs.\ therein): 
\begin{equation}
\lambda_i = \frac{d_i}{\pi} \frac{\Gamma_i}{\omega_i^2} \frac{2}{N(E_F)},
\end{equation}
where $d_i$ is the degeneracy of the mode. By assuming a density of states 
at the Fermi level $N(E_F) \sim 5$ states/eV, as deduced from ESR, and using 
the fitted parameters in Tab.~\ref{tab:Raman_data}, we obtain the relatively 
high $\lambda_{{\mathrm{H}_g}(1)} = 1.4$ and $\lambda_{{\mathrm{H}_g}(2)} = 0.6$,
whereas the contribution from the A$_g$ modes is completely negligible. 

In conclusion, we have shown that an A$_{4}$C$_{60}$ fulleride 
which assumes an fcc structure is metallic. Although there are 
already some indications that A$_{4}$C$_{60}$ fullerides are 
on the verge of a metal-to-insulator transition (MIT) (
Rb$_{4}$C$_{60}$ at high pressure \cite{Kerkoud96}, tetragonal 
Na$_{4}$C$_{60}$ 
\cite{Oszlanyi98}), the metallicity is not a direct 
consequence of the lattice contraction due to the small dimension of 
the intercalated ions. Indeed, K$_{4}$C$_{60}$ presents the same inter-ball 
distance as Li$_{4}$C$_{60}$, but it is an insulator.
This indicates that lattice symmetry plays a dominant role 
\cite{Han00}, unlike molecular effects such as Jahn-Teller distortions,
which tend to stabilize an insulator \cite{Capone00}. 
Compared to superconducting A$_{3}$C$_{60}$, this system has a broader 
bandwidth (probably related to the presence of asymmetric Li ions), but a 
relatively stronger electron-phonon coupling. This suggests that a 
moderate lattice expansion (achieved, for example, by the co-intercalation 
of small molecules like ammonia \cite{Ricco06}) would preserve the metallic state, 
while inhibiting the formation of a polymer at low temperatures. 
Such a new material is very likely to show high-$T_c$ superconductivity.

We thank Irene Margiolaki at ESRF for technical assistance and 
acknowledge financial support from the EC FP6-NEST ``Ferrocarbon'' 
project.


\end{document}